\documentclass[conference]{IEEEtran}
\IEEEoverridecommandlockouts
\usepackage{cite}
\usepackage{overpic}
\usepackage{amsmath,amssymb,amsfonts}
\usepackage{graphicx}
\usepackage{textcomp}
\usepackage{xcolor}
\usepackage{float}
\usepackage{amsthm}
\usepackage{graphicx}
\usepackage{epstopdf}
\usepackage{amsmath,bm}
\usepackage{amsfonts}
\usepackage{amssymb}
\usepackage{color}
\usepackage{multirow}
\usepackage{multicol}
\usepackage{soul,xcolor}
\usepackage{algorithm}
\usepackage{algpseudocode}

\theoremstyle{plain}

\newtheorem{lemma}{Lemma}

\newcommand{\vect}[1]{\mathbf{#1}}

\def\tr{\mathrm{tr}}
\def\rank{\mathrm{rank}}
\def\Htran{\mbox{\tiny $\mathrm{H}$}}
\def\Ttran{\mbox{\tiny $\mathrm{T}$}}
\def\CN{\mathcal{N}_{\mathbb{C}}} 
\def\imagunit{\mathsf{j}} 
\def\m{\rm}

\def\sinc{\mathrm{sinc}}

\begin{document}

\title{\huge Channel Modeling and Channel Estimation for \\ Holographic Massive MIMO with Planar Arrays\vspace{-0.4cm}
}

\author{\"Ozlem Tu\u{g}fe Demir, \emph{Member, IEEE}, Emil Bj{\"o}rnson, \emph{Fellow, IEEE}, Luca Sanguinetti, \emph{Senior Member, IEEE}\vspace{-2cm}
\thanks{\"O. T. Demir and E.~Bj\"ornson are with the KTH Royal Institute of Technology, 16440 Kista, Sweden (\{ozlemtd, emilbjo\}@kth.se). \newline \indent L.~Sanguinetti is with the University of Pisa, Dipartimento di Ingegneria dell'Informazione, 56122 Pisa, Italy (luca.sanguinetti@unipi.it). \newline 
}\vspace{-3cm}
}

\maketitle
\begin{abstract}
In a realistic wireless environment, the multi-antenna channel usually exhibits spatially correlation fading. This is more emphasized when a large number of antennas is densely deployed, known as holographic massive MIMO (multiple-input multiple-output). In the first part of this letter, we develop a channel model for holographic massive MIMO by considering both non-isotropic scattering and directive antennas. With a large number of antennas, it is difficult to obtain full knowledge of the spatial correlation matrix. In this case, channel estimation is conventionally done using the least-squares (LS) estimator that requires no prior information of the channel statistics or array geometry. In the second part of this letter, we propose a novel channel estimation scheme that exploits the array geometry to identify a subspace of reduced rank that covers the eigenspace of any spatial correlation matrix. The proposed estimator outperforms the LS estimator, without using any user-specific channel statistics.
\end{abstract}

\vspace{-2mm}
\section{Introduction}
\vspace{-1mm}

The base stations (BSs) in 5G are equipped with a large number of antennas to enable efficient beamforming and spatial multiplexing of user equipments (UEs) \cite{Bjornson2019d}. 
This is known as massive MIMO (multiple-input multiple-output)~\cite{massivemimobook} and the spectral efficiency increases with the number of antennas. The asymptotic performance limits have received much attention in the literature, under the assumption that the array aperture grows large. However, in practice, the array aperture is limited. Hence, the corresponding asymptotic limit is a spatially-continuous aperture with densely deployed antennas, known as \emph{holographic MIMO} \cite{huang2020holographic,pizzo2020spatially}
and large intelligent surface \cite{hu2018beyond}. We will use the term \emph{holographic massive MIMO} since it is a natural extension of current massive MIMO technology. By deploying more antennas in a given area, one can spatially multiplex greater number of users, reduce interference, and increase the beamforming gain \cite{Bjornson2019d,hu2018beyond,pizzo2020asilomar}. 

One way to approximately realize a spatially-continuous aperture is to use a rectangular surface with  inter-antenna spacing far less than half of the wavelength \cite{Bjornson2019d} and there exist several candidate hardware implementations \cite{huang2020holographic}. One of the important features of such a rectangular surface is the inevitable spatial correlation among the channel realizations of the BS antennas. This leads to a rank-deficient spatial correlation matrix \cite{Bjornson2021b}. The ratio of the spatial correlation matrix rank to the channel dimension decreases as the antennas are deployed more densely. Hence, this low-rank feature (spatial correlation) is more emphasized for holographic massive MIMO channels. To perform minimum mean-squared error (MMSE) channel estimation, 
the spatial correlation matrix of a UE should be known at the BS. In a holographic massive MIMO system with thousands of antennas, it is challenging to both acquire the spatial correlation matrix and implement the MMSE estimator. An alternative is to use the least-squares (LS) estimator, which utilizes no prior information. However, the performance of LS to MMSE is inferior at low values of the signal-to-noise ratio (SNR). Moreover, the LS scheme neglects the spatial correlation induced by the array geometry, which is naturally known in a given deployment. In this letter, we propose a novel channel estimation scheme that exploits the part of the correlation created by the array geometry to outperform the LS estimator. We motivate our scheme analytically and show that it is guaranteed to outperform the LS estimator. The performance gap increases as the rank deficiency induced by the array geometry increases.

To study the performance of the proposed channel estimator, we need a spatial correlation model for holographic massive MIMO channels. In \cite{pizzo2020spatially}, a small-scale fading model for the holographic MIMO channel is analyzed from an electromagnetic perspective. In this letter, we develop channel modeling by providing an exact integral expression for the  spatial correlation matrix with non-isotropic scattering and directive antennas different from \cite{pizzo2020spatially}. We derive a closed-form expression suitable for large-scale performance evaluation. In this way, we develop further insights based on simulations.

\section{System and Channel Modeling}

We consider an uplink holographic massive MIMO system where the BS is equipped with a uniform planar array (UPA) with $M$ antennas. Following \cite[Fig.~1]{Bjornson2021b}, the number of antennas per row and per column are denoted by $M_{\rm H}$ and $M_{\rm V}$, respectively, and $M=M_{\rm H}M_{\rm V}$.
The horizontal and vertical antenna spacing is $\Delta$.
We are particularly targeting use cases with thousands of antennas and antenna spacing below half of the wavelength $\lambda$. 
The antennas are indexed row-by-row by $m\in[1,M]$, thus the location of the $m$th antenna with respect to the origin in \cite[Fig.~1]{Bjornson2021b} is $\vect{u}_m = [ 0, \, \,\, i(m) \Delta,  \,\,\, j(m) \Delta]^{\Ttran}$
where $i(m) =\mathrm{mod}(m-1,M_{\m H})$ and $j(m) =\left\lfloor(m-1)/M_{\m H}\right\rfloor$
are the horizontal and vertical indices of element $m$, respectively. Notice that $\mathrm{mod}(\cdot,\cdot)$ denotes the modulus operation and $\lfloor \cdot \rfloor$ truncates the argument. Using this notation, if a plane wave is impinging on the UPA from the azimuth angle $\varphi$ and elevation angle $\theta$, the array response vector is~\cite[Sec.~7.3]{massivemimobook}

\begin{equation}\label{eq:array-response}
\vect{a}(\varphi,\theta) = \left[e^{\imagunit\vect{k}(\varphi,\theta)^{\Ttran}\vect{u}_1},\dots,e^{\imagunit\vect{k}(\varphi,\theta)^{\Ttran}\vect{u}_M}\right]^{\Ttran}
\end{equation}
where $\vect{k}(\varphi, \theta) = \frac{2\pi}{\lambda}\left[\cos(\theta) \cos(\varphi), \,\,\, \cos(\theta) \sin(\varphi), \,\,\, \sin(\theta)\right]^{\Ttran}$ is the wave vector.

We consider an arbitrary single-antenna UE and denote its channel to the BS by $\vect{h}\in \mathbb{C}^M$. 
When the UE transmits, the received signal at the BS will generally consist of a superposition of multipath components that can be expanded as a continuum of plane waves \cite{Sayeed2002a}. This is also a tight approximation for spherical waves, beyond Fresnel distance.
Hence, we may write
\begin{equation} \label{eq:channel1}
\vect{h} =   \iint_{-\pi/2}^{\pi/2} g(\varphi,\theta) \vect{a}(\varphi,\theta) d\theta d\varphi 
\end{equation}
where the \emph{angular spreading function} $g(\varphi,\theta)$ specifies the gain and phase-shift from each direction $(\varphi,\theta)$ and $ \vect{a}(\varphi,\theta)$ is the array response vector in  \eqref{eq:array-response} for $\varphi \in [-\frac{\pi}{2},\frac{\pi}{2}]$ and $\theta \in [-\frac{\pi}{2},\frac{\pi}{2}]$.
Note that the waves only arrive from directions in front of the array; that is, $\varphi \in [-\frac{\pi}{2},\frac{\pi}{2}]$. While the analysis applies also to the radiative near-field, we focus on far-field channels  for which the wavefronts are approximately plane over the array. 

The microscopic fading created by small-scale mobility is captured by $g(\varphi,\theta)$ being a time-varying variable that can be modeled stochastically. We consider the conventional block fading model, where the channel $\vect{h}$ is constant within one time-frequency block and takes independent realization across blocks from a stationary stochastic distribution.
In accordance to \cite{Sayeed2002a}, we model $g(\varphi,\theta)$ as a spatially uncorrelated circularly symmetric Gaussian stochastic process with cross-correlation
\begin{equation} \label{eq:scattering-correlation-model}
\mathbb{E} \{ g(\varphi,\theta) g^*(\varphi',\theta') \} = \beta f(\varphi,\theta) \delta(\varphi-\varphi')  \delta(\theta-\theta')
\end{equation}
where $\delta(\cdot)$ denotes the Dirac delta function, $\beta$ denotes the average channel gain (i.e., capturing pathloss and shadowing), and $f(\varphi,\theta)$ is the normalized \emph{spatial scattering function} \cite{Sayeed2002a}.  The latter function describes the angular multipath distribution and the directivity gain of the antennas, and it is normalized so that $\iint f(\varphi,\theta) d\theta d\varphi  = 1$. It thus follows that
\begin{equation} \label{eq:corr-Rayleigh}
\vect{h} \sim \CN(\vect{0},\vect{R})
\end{equation}
which is a correlated Rayleigh fading channel fully characterized by the spatial correlation matrix
\begin{equation} \label{eq:spatial-correlation}
\vect{R} = \mathbb{E}\{ \vect{h} \vect{h}^{\Htran} \} = \beta  \iint_{-\pi/2}^{\pi/2} f(\varphi,\theta) \vect{a}(\varphi,\theta) \vect{a}^{\Htran}(\varphi,\theta) d\theta d\varphi 
\end{equation}
where the last equality follows from \eqref{eq:scattering-correlation-model}. Notice that $\tr(\vect{R}) = M\beta$.
By utilizing the structure of the array response in \eqref{eq:array-response}, we obtain the following general channel model.

\begin{lemma}
For any given spatial scattering function $f(\varphi,\theta)$, the channel vector $\vect{h} \sim \CN(\vect{0},\vect{R})$ of the UPA has the spatial correlation matrix $\vect{R}$ with
the $(m,l)$th entry given by
\begin{align}  \label{eq:spatial-correlation2}
\left[\vect{R}\right]_{m,l} = &\beta  \!
 \iint_{-\pi/2}^{\pi/2} f(\varphi,\theta) \nonumber\\
 &\hspace{12mm} \times e^{\imagunit2\pi \left( d_{{\rm H}}^{ml}\sin(\varphi)\cos(\theta) + d_{{\rm V}}^{ml}\sin(\theta) \right)} d\theta d\varphi 
\end{align}
where the horizontal and vertical distances between antenna $m$ and $l$ (normalized by the wavelength) are given by
\begin{align}
d_{{\rm H}}^{ml} = \frac{ \left( i(m) - i(l) \right) \Delta}{\lambda}, \quad d_{{\rm V}}^{ml} = \frac{ \left( j(m) - j(l) \right) \Delta}{\lambda}.
\end{align}
\end{lemma}

The double-integral in \eqref{eq:spatial-correlation2} can be computed numerically for any spatial scattering function, but some functions also lead to closed-form expressions. One example is an \emph{isotropic scattering environment} where the multipath components are equally strong in all directions and the antennas are isotropic (i.e., $f(\varphi,\theta) = \cos(\theta) / (2\pi)$, where the cosine comes from the spherical coordinate system). We denote the resulting correlation matrix by $\vect{R}_{\rm iso}$ and the $(m,l)$th entry is~\cite{Bjornson2021b}
\begin{equation}\label{R-iso}
    \left[\vect{R}_{\rm iso} \right]_{m,l} = \beta \sinc \left( 2 \sqrt{\left(d_{\rm H}^{ml}\right)^2+\left(d_{\rm V}^{ml}\right)^2}\right)
\end{equation}
where $\sinc(x) = \sin(\pi x)/ (\pi x)$ is the sinc function. 
The expression in \eqref{R-iso} shows that two antennas that are spaced apart by an integer multiple of $\lambda/2$ will exhibit mutually uncorrelated fading. However, this can never be satisfied for all pairs of antennas in a UPA \cite{Bjornson2021b}. Therefore, such an array will always exhibit spatially correlated fading.

\subsection{Clustered Scattering Model with Directive Antennas}

We will now develop a spatial correlation matrix model for a realistic scenario with directive antennas and a non-isotropic scattering environment.
In particular, the scattered waves from a UE reach the BS from a set of
$N$ angular clusters, which could represent different objects in the environment.
This is a generalization of \cite[Sec.~VIII]{Demir2021RIS}, where a  UPA with a single cluster and isotropic antennas is considered. We consider antennas with a cosine directivity pattern along the azimuth and elevation angles given by \cite{kumar20192d}
\begin{align} \label{eq:antenna-gain}
    \!\!\!\mathcal{D}(\varphi,\theta)  \propto\cos^a(\varphi)\cos^b(\theta),\,\varphi \in \left[-\frac{\pi}{2},\frac{\pi}{2}\right], \theta \in \left[-\frac{\pi}{2},\frac{\pi}{2}\right]\!\!
\end{align}
where the exponents $a\geq 0$ and $b\geq 0$ determine the directivity. Larger values result in narrower patterns.  The proportionality constant should make
$\iint \mathcal{D}(\varphi,\theta) \cos(\theta) d\varphi d\theta = 4\pi$.

We assume that cluster $n$ is centered around the nominal azimuth and elevation angles $\varphi_n$ and $\theta_n$, for $n=1,\ldots,N$. Let $\delta_n$ and $\epsilon_n$ denote the respective angular deviations. Using an independent von Mises distribution in the considered angle range for the multipath components around the nominal angles  \cite[Eq. (10)]{abdi2002space} and including the directivity pattern from \eqref{eq:antenna-gain}, we obtain the spatial scattering function for cluster $n$ as
\begin{align} \label{eq:spatial-scattering-cluster}
f_n(\delta_n,\epsilon_n)=& \mathcal{A} \mathcal{P}_n \cos^a(\varphi_n+\delta_n)\cos^{b+1}(\theta_n+\epsilon_n) \nonumber\\
&\hspace{-16mm}\times e^{\frac{\cos(2\delta_n)}{4\sigma_{\varphi}^2}}e^{\frac{\cos(2\epsilon_n)}{4\sigma_{\theta}^2}}, \ \left\vert \varphi_n+\delta_n\right\vert\leq \frac{\pi}{2}, \ \left\vert\theta_n+\epsilon_n\right\vert\leq \frac{\pi}{2}
\end{align}
where $\mathcal{P}_n\geq 0$ is the normalized power of cluster $n$ and the scalar $\mathcal{A}>0$
must be selected so that 
$\sum_{n=1}^N\iint f_n(\delta_n,\epsilon_n) d\delta_n d\epsilon_n = 1$. The additional $\cos(\theta_n+\epsilon_n)$ term comes from the width of a solid angle in the spherical coordinate system (which also exists in the isotropic antenna pattern from the differential of the solid angle). As the angular standard deviations $\sigma_{\varphi}$ and $\sigma_{\theta}$ go to zero, the von Mises distribution approaches the Gaussian distribution. Using   \eqref{eq:spatial-correlation2} and \eqref{eq:spatial-scattering-cluster}, the $(m,l)$th entry of $\vect{R}$ becomes
\begin{align} \label{eq:exact-R}
\left[\vect{R}\right]_{m,l} = & \mathcal{A}\beta \sum_{n=1}^N \mathcal{P}_n
\int_{-\theta_n-\pi/2}^{-\theta_n+\pi/2}\int_{-\varphi_n-\pi/2}^{-\varphi_n+\pi/2}  e^{\imagunit2\pi d_{{\rm V}}^{ml}\sin(\theta_n+\epsilon_n)} \nonumber \\
& \times e^{\imagunit2\pi d_{{\rm H}}^{ml}\sin(\varphi_n+\delta_n)\cos(\theta_n+\epsilon_n)} \cos^a(\varphi_n+\delta_n)\nonumber\\
&\times \cos^{b+1}(\theta_n+\epsilon_n)e^{\frac{\cos(2\delta_n)}{4\sigma_{\varphi}^2}}e^{\frac{\cos(2\epsilon_n)}{4\sigma_{\theta}^2}}d\delta_n d\epsilon_n.
\end{align}
 The integrals in \eqref{eq:exact-R} can be computed numerically but that is computationally demanding for large arrays since $\bf R$ has $M^2$ entries.
The following lemma provides a closed-form approximation 
that is tight for narrow angular clusters. The proof is omitted due to space limitations.

\begin{lemma}\label{lemma:derivation-spatial-correlation}
When the angular deviations are small in the sense that $\cos(\delta_n)\approx 1$, $\cos(\epsilon_n)\approx 1$, $\sin(\delta_n)\approx \delta_n$, $\sin(\epsilon_n)\approx \epsilon_n$, for $n=1,\ldots,N$, then the $(m,l)$th entry of the spatial correlation matrix can be tightly approximated as
\begin{align}
&\left[\vect{R}\right]_{m,l}  \approx \frac{\beta}{\sum_{n=1}^N\mathcal{P}_n\cos^a\left(\varphi_n\right)\cos^{b+1}\left(\theta_n\right)} \nonumber\\
&\times \sum_{n=1}^N \mathcal{P}_n  \frac{ A_{mln} \widetilde{\sigma}_{mln}}{\sigma_{\varphi}}e^{-\frac{B_{mln}^2\widetilde{\sigma}^2_{mln}}{2}} \nonumber\\
&\times\! e^{\frac{D_{mln}^2\sigma_{\theta}^2\left(C_{mln}^2\sigma_{\theta}^2\widetilde{\sigma}_{mln}^2\!-1\right)}{2}}  e^{-\imagunit B_{mln}C_{mln}D_{mln}\sigma_{\theta}^2\widetilde{\sigma}_{mln}^2 } \nonumber \\
&\times\! \left(X_{mln}-Y_{mln}\left(\imagunit B_{mln}\widetilde{\sigma}_{mln}^2-C_{mln}D_{mln}\sigma_{\theta}^2\widetilde{\sigma}_{mln}^2\right)\right) 
\end{align}
with
$A_{mln}= e^{\imagunit2\pi d_{{\rm H}}^{ml}\sin(\varphi_n)\cos(\theta_n)}e^{\imagunit2\pi d_{{\rm V}}^{ml}\sin(\theta_n)}$, $B_{mln}=  2\pi d_{{\rm H}}^{ml}\cos(\varphi_n)\cos(\theta_n)$,
    ${C_{mln}= -2\pi d_{{\rm H}}^{ml}\cos(\varphi_n)\sin(\theta_n)}$,
    ${D_{mln}=-2\pi d_{{\rm H}}^{ml}\sin(\varphi_n)\sin(\theta_n)+2\pi d_{{\rm V}}^{ml}\cos(\theta_n)}$, ${\widetilde{\sigma}^2_{mln}= \frac{\sigma_{\varphi}^2}{1+C_{mln}^2\sigma_{\varphi}^2\sigma_{\theta}^2}}$, $X_{mln}= \cos^a(\varphi_n)\big(\cos^{b+1}(\theta_n)-\imagunit (b+1)\cos^{b}(\theta_n)\sin(\theta_n) \sigma_{\theta}^2D_{mln}\big)$, $Y_{mln}= a\cos^{a-1}(\varphi_n)\sin(\varphi_n)\cos^{b+1}(\theta_n)+\imagunit \sigma_{\theta}^2(b+1)\cos^{b}(\theta_n)$ $\times\sin(\theta_n)\big( \cos^a(\varphi_n)C_{mln}-a\cos^{a-1}(\varphi_n)\sin(\varphi_n)D_{mln}\big)$.

\end{lemma}

\begin{figure}[t!]\vspace{-0.5cm}
\hspace{-3mm}
			\includegraphics[trim={0cm 0.cm 0cm 0cm},clip,width=3.6in]{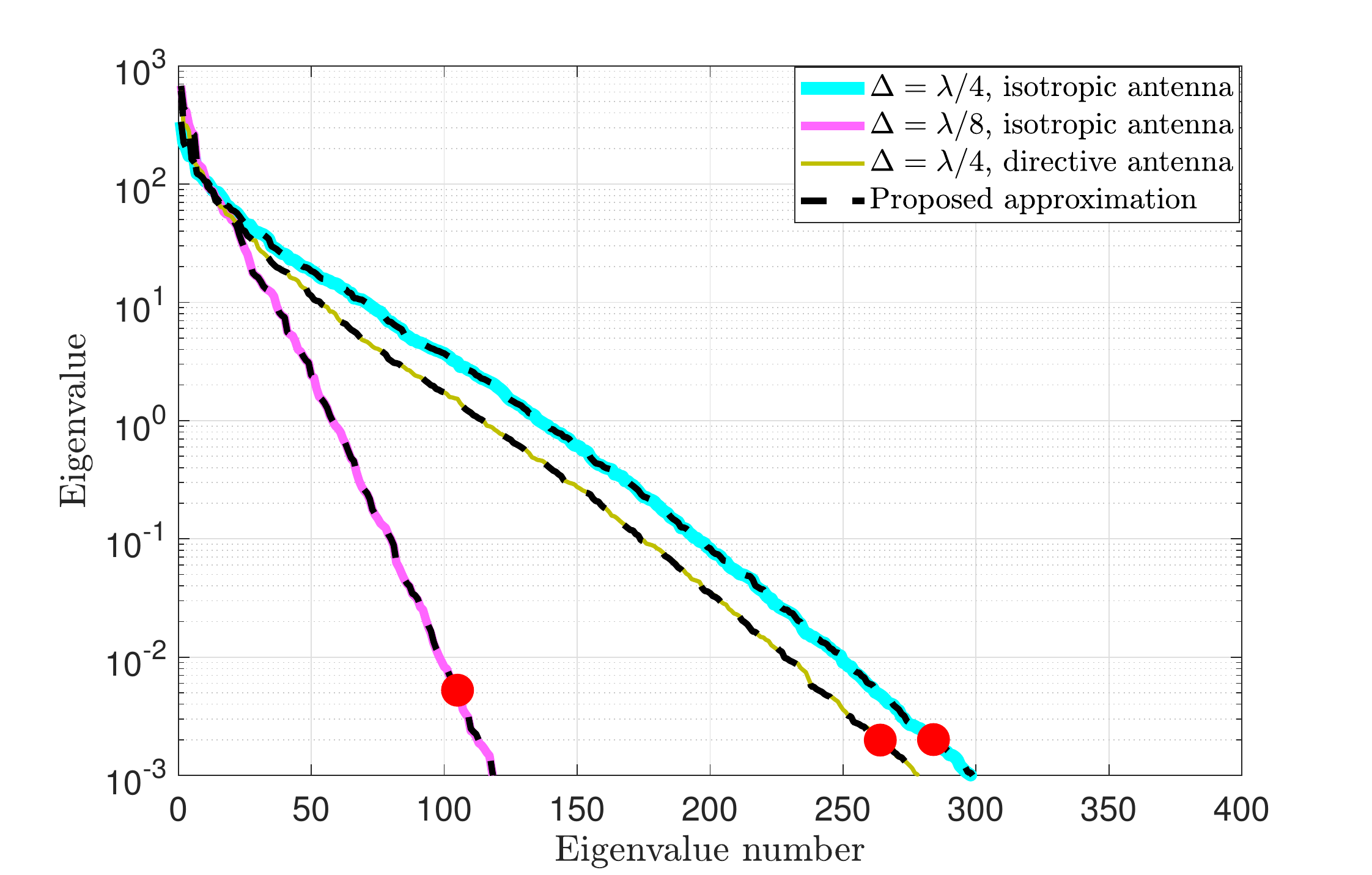}
			\vspace{-0.75cm}
			\caption{Sorted eigenvalues for the exact and approximate spatial correlation matrices. A $64\times 64$ UPA with varying antenna patterns and spacings is used.} \label{fig:eigenvalue} \vspace{-6mm}
\end{figure}

In Fig.~\ref{fig:eigenvalue}, the tightness of the approximate closed-form expression in Lemma~\ref{lemma:derivation-spatial-correlation} is demonstrated. The eigenvalues of the exact (obtained through many hours of numerical integration) and approximate closed-form spatial correlation matrices are plotted in decreasing order. We consider a $64\times64 $ UPA with the antennas having either an isotropic or a cosine pattern proportional to $\cos^5(\varphi)\cos^5(\theta)$. There are $N=20$ clusters with exponential power delay profile, which are generated according to the urban macrocell environment model in sub-6 GHz frequency band by following \cite[p.~54]{series2017guidelines}. The BS and UE  are 25\,m and 1.5\,m above the ground, respectively. The nominal azimuth and elevation angles are determined from the cluster powers according to \cite[p.~55-58]{series2017guidelines}. The per-cluster angular standard deviations are $\sigma_{\varphi} = \sigma_{\theta} = 2^{\circ}$. The filled circles in Fig.~\ref{fig:eigenvalue} show the effective rank, containing a fraction $1-10^{-5}$ of the sum of all eigenvalues. Fig.~\ref{fig:eigenvalue} shows that the eigenvalues match very closely. Moreover, the correlation matrix distance used in \cite[Fig.~3]{Bjornson2021b} between the exact and approximate spatial correlation matrices (it is between 0 and 1) is approximately $1.75\cdot10^{-5}$ and $10^{-5}$ for isotropic antenna case with $\Delta=\lambda/4$ and $\Delta=\lambda/8$, respectively. For the case of directive antennas with the cosine pattern and $\Delta=\lambda/4$, it is $0.9\cdot10^{-5}$. This demonstrates the tightness of the approximate closed-form expression in Lemma~\ref{lemma:derivation-spatial-correlation}. 

The rank of the spatial correlation matrix is reduced when having directive antennas, but the reduction in antenna spacing has a sharper impact on the rank than using directive antennas.

\vspace{-1mm}
\section{Channel Estimation} 

The BS must estimate $\vect{h}$ in each coherence block to use the $M$ antennas for coherent beamforming. The standard approach is that the UE sends a predefined pilot sequence. From \cite[Sec.~3]{massivemimobook}, the received signal at the BS is\footnote{When multiple UEs send orthogonal pilots, the channel estimation is done separately by despreading the received signal at the BS. The received signal denotes the despreaded signal for a typical UE.} 
\begin{align} \label{eq:y}
    \vect{y} = \sqrt{\rho}\vect{h} + \vect{n}
\end{align}
where $\rho>0$ is the pilot SNR and $\vect{n}\sim \CN\left(\vect{0}, \vect{I}_M \right)$.

Different channel estimation schemes can be utilized at the BS, depending on the available statistical information.
If the full matrix $\vect{R}$ in \eqref{eq:spatial-correlation} is known, the MMSE estimate of $\vect{h}$ is
\begin{align}\label{eq:MMSE-estimate}
\widehat{\vect{h}}_{\rm MMSE} = \sqrt{\rho}\vect{R}\left(\rho\vect{R}+\vect{I}_M\right)^{-1}\vect{y}.
\end{align}
However, since $\vect{R}$ contains $M^2$ entries, it is challenging to acquire it in practice, particularly when $M$ is large and/or when the UE is only transmitting a small data packet.

An alternative approach is the LS estimator that only requires knowledge of the pilot SNR, $\rho$, and provides
\begin{align} \label{eq:LS-estimate}
   \widehat{ \vect{h}}_{\rm LS} = \frac{\vect{y}}{\sqrt{\rho}}.
    \end{align}
We will show that this estimator is unnecessarily conservative.

\subsection{Subspace-Based Channel Estimation}

The rank of the spatial correlation matrix $\vect{R}_{\rm iso}$  for isotropic scattering and antennas is approximately $\pi M (\Delta/\lambda)^2$ when $M$ is large and $\Delta$ is small \cite[Prop.~2]{Bjornson2021b}. For example, for $\Delta= \lambda/2$, the rank is $M \pi/4$, thus 21\% of the eigenvalues are zero.
The rank-deficiency grows when the antenna spacing reduces due to the resulting spatial oversampling;  80\% of the eigenvalues are zero when using $\Delta= \lambda/4$.
With non-isotropic scattering and/or non-isotropic antennas, we might obtain even stronger rank-deficiency.
However, since the low rank of $\vect{R}_{\rm iso}$ is caused purely by spatial oversampling, an interesting question is: \emph{Can we utilize the array geometry to improve the channel estimation when the true correlation matrix $\vect{R}$ is unknown?}

In this subsection, we will develop a subspace-based estimation method that outperforms LS estimation without requiring any UE-dependent prior information but only the array geometry.
We will first show how the MMSE estimator implicitly exploits low-rank correlation matrices.
We let $1\leq r \leq M$ denote the rank of $\vect{R}$, i.e., $\rank\left(\vect{R}\right)=r$. The \emph{compact} eigenvalue decomposition is denoted as $\vect{R}=\vect{U}_1\vect{\Lambda}_1\vect{U}_1^{\Htran}$, where the diagonal matrix $\vect{\Lambda}_1 \in \mathbb{C}^{r \times r}$ contains the non-zero eigenvalues and the columns of $\vect{U}_1\in \mathbb{C}^{M \times r}$ contains the corresponding orthonormal eigenvectors.

 The channel $\vect{h}$ can be expressed as $\vect{h} = \vect{R}^{\frac12}\vect{v} = \vect{U}_1\vect{\Lambda}_1^{\frac12}\vect{v}$,
where $\vect{v} \sim \CN(\vect{0},\vect{I}_r)$. Hence, all channel realizations exist in the subspace spanned by $\vect{U}_1$ (i.e., they are
linear combinations of its columns).

The MMSE estimate $\widehat{\vect{h}}_{\rm MMSE}$ in \eqref{eq:MMSE-estimate} can be expressed as 
\begin{align}\label{eq:MMSE-estimate2}
\widehat{\vect{h}}_{\rm MMSE} =  \vect{U}_1 \left( \frac{1}{\sqrt{\rho}} \vect{D}_1 \right) \vect{U}_1^{\Htran}\vect{y} 
\end{align}
where $\vect{D}_1 = \rho \vect{\Lambda}_1 ( \rho \vect{\Lambda}_1 + \vect{I}_r)^{-1}$ is a diagonal matrix.
Hence, the MMSE estimator in \eqref{eq:MMSE-estimate2} carries out three operations: 1) $ \vect{U}_1^{\Htran}\vect{y} $ projects the received signal onto the subspace spanned by $\vect{U}_1$; 2) The resulting $r$ channel dimensions are MMSE estimated using the scaling factors in $\vect{D}_1 /\sqrt{\rho}$;  3) The estimate is brought back to the original $M$-dimensional space using $ \vect{U}_1$.

If only the subspace spanned by $\vect{U}_1$ is known, not the eigenvalues necessary to compute $\vect{D}_1$, we can replace $\vect{D}_1$ in \eqref{eq:MMSE-estimate2} by the corresponding LS estimator. We call this the \emph{reduced-subspace LS (RS-LS)} estimator:
\begin{align} \label{eq:RS-LS-estimate}
    \widehat{\vect{h}}_{\rm RS-LS} = \frac{\vect{U}_1\vect{U}_1^{\Htran}\vect{y}}{\sqrt{\rho}}.
\end{align}
We notice that RS-LS is obtained from the MMSE estimator by replacing $\vect{D}_1$ with $\vect{I}_r$. Moreover,
we notice that $\vect{D}_1 \to \vect{I}_r$ as $\rho \to \infty$, thus we expect RS-LS to perform similarly as MMSE at high SNR, but to provide larger estimation errors in other situations. In any case, the subspace projection in RS-LS removes the noise from $M-r$ dimensions, effectively increasing the SNR by a factor $M/r$ in the estimation phase.

To dispense from knowledge of $\vect{R}$, we can instead utilize RS-LS along with some other correlation matrix $\overline{\vect{R}}$ that is not representing a particular UE but the general array geometry.
For example, we can set $\overline{\vect{R}}=\vect{R}_{\rm iso}$ to ensure that all $\rank(\vect{R}_{\rm iso})$ possible channel dimensions are considered by the estimator. This property can be formalized as follows.

\begin{lemma} \label{lemma:span} Let $\overline{\vect{R}}$ and $\vect{R}$ be two spatial correlation matrices obtained using the same array geometry. The spatial scattering functions corresponding to these correlation matrices are denoted by $\overline{f}(\varphi,\theta)$  and $f(\varphi,\theta)$, respectively, for $\varphi\in[-\pi/2,\pi/2]$ and $\theta\in[-\pi/2,\pi/2]$. Assume that $\overline{f}(\varphi,\theta)$  and $f(\varphi,\theta)$ are either continuous at each point on its domain or contain Dirac delta functions.

If the domain of  $\overline{f}(\varphi,\theta)$ for which $\overline{f}(\varphi,\theta)>0$ contains the domain $f(\varphi,\theta)$ for which $f(\varphi,\theta)>0$, then the subspace spanned by the columns of $\overline{\vect{R}}$ contains the subspace spanned by the columns of $\vect{R}$.
\begin{proof}
The proof is given in the Appendix.
\end{proof}
\end{lemma}

An important special case of  Lemma~\ref{lemma:span} is obtained for $\overline{\vect{R}}=\vect{R}_{\rm iso}$ by noting that the span of the correlation matrix with isotropic scattering spans the entire angular domain. Hence, for a given array geometry, the subspace spanned by the columns of $\vect{R}_{\rm iso}$ contains the subspace spanned by any other $\vect{R}$.

For a given  $\overline{\vect{R}}$ that satisfies the condition in Lemma~\ref{lemma:span}, the resulting so-called \emph{conservative RS-LS} estimator is
\begin{align} \label{eq:RS-LS-estimate-approx}
    \widehat{\vect{h}}_{\rm RS-LS}^{\rm conserv} = \frac{\overline{\vect{U}}_1\overline{\vect{U}}_1^{\Htran}\vect{y}}{\sqrt{\rho}}
\end{align}
where the columns $\overline{\vect{U}}_1\in \mathbb{C}^{M \times \overline{r}}$ 
are the orthonormal eigenvectors corresponding to the  $ \overline{r}$ non-zero eigenvalues of $\overline{\vect{R}}$. 

Although we have focused on the channel estimation for a typical UE, we note that $\vect{R}_{\rm iso}$ or any other $\overline{\vect{R}}$ that satisfies the condition in Lemma~\ref{lemma:span} is common among all UEs and the proposed channel estimation can be applied for any UE in a multi-UE setup irrespective of their correlation matrices.

 \begin{figure}[t!] \vspace{-0.5cm}
\hspace{-3mm}
	\includegraphics[trim={0cm 0.cm 0cm 0cm},clip,width=3.6in]{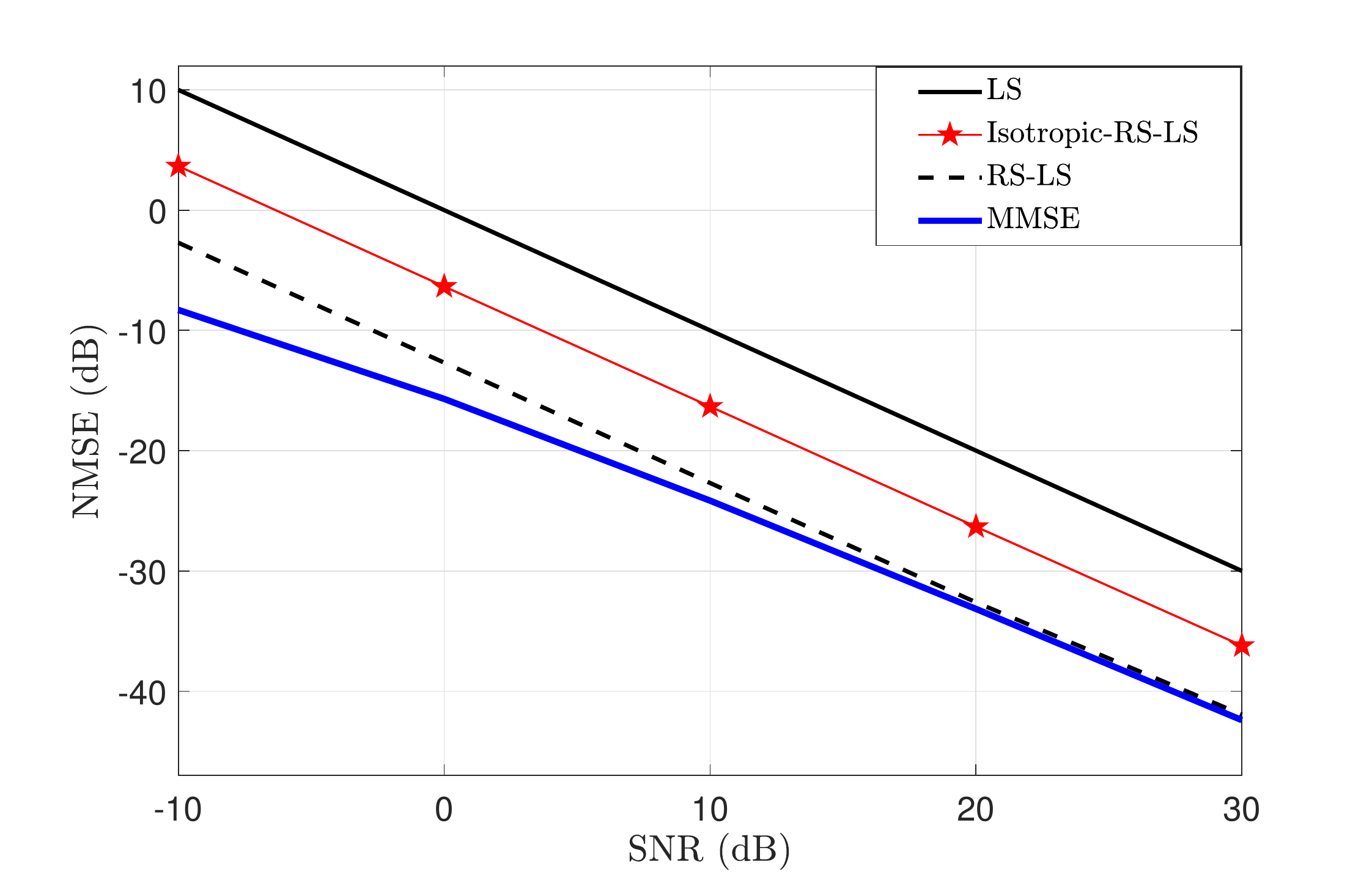}
			\vspace{-0.75cm}
			\caption{NMSE versus SNR for the $128\times128$ UPA  with $\Delta=\lambda/4$.} \label{fig:M128d25} \vspace{-5mm}
\end{figure}

 \section{Comparison of Channel Estimation Schemes}

We will now quantify the channel estimation performance of the considered schemes numerically in terms of the normalized mean-square error (NMSE). We use the same cluster and angular deviation properties as in Fig.~\ref{fig:eigenvalue} with cosine antenna pattern proportional to $\cos(\varphi)\cos(\theta)$ and plot the NMSE versus SNR, $\rho$, with $\beta=1$. The MMSE, LS, RS-LS, and Isotropic-RS-LS are obtained from \eqref{eq:MMSE-estimate}, \eqref{eq:LS-estimate}, \eqref{eq:RS-LS-estimate}, and  \eqref{eq:RS-LS-estimate-approx} with $\vect{R}_{\rm iso}$, respectively. 

The antenna array is $128\times 128$ UPA and the antenna spacing is $\Delta=\lambda/4$ in 
Fig.~\ref{fig:M128d25}. The optimal MMSE estimator provides the lowest NMSE, while the conventional statistics-unaware LS estimator provides a 12\,dB higher NMSE.
The proposed RS-LS estimator applies LS within the subspace spanned by the true spatial correlation matrix, thereby eliminating noise from the nullspace of the correlation matrix.
As the SNR increases, the gap between RS-LS and MMSE vanishes. However, RS-LS still uses the true correlation matrix. The proposed estimator ``Isotropic-RS-LS'' is not utilizing such information but only the array geometry.
Isotropic-RS-LS provides a 6\,dB performance gain over LS, but there is a gap to RS-LS. The reason is that the true correlation matrix has a lower rank  than $\vect{R}_{\rm iso}$ (the effective ranks are 881 vs.~3808).

\begin{figure}[t!]
\vspace{-0.6cm}
\hspace{-3mm}
			\includegraphics[trim={0cm 0.cm 0cm 0cm},clip,width=3.6in]{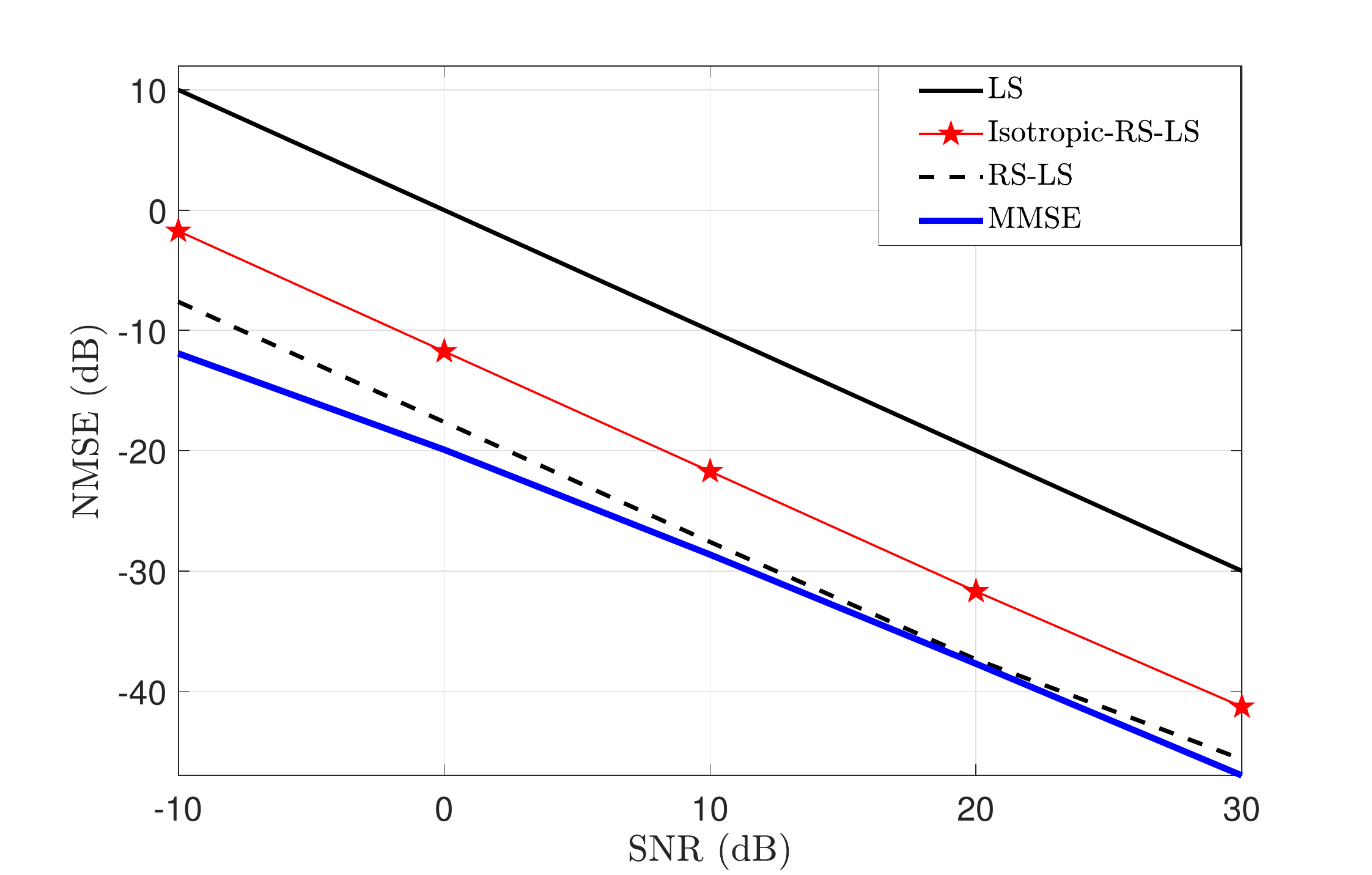}
			\vspace{-0.75cm}
			\caption{NMSE versus SNR for the $128\times128$ UPA  with $\Delta=\lambda/8$.} \label{fig:M128d125} \vspace{-5mm}
\end{figure}

 We reduce the antenna spacing to
 $\Delta=\lambda/8$ in Fig.~\ref{fig:M128d125}. Since the array is denser, the ratio of the rank of the spatial correlation matrix to $M$ is smaller and the NMSE gap between Isotropic-RS-LS and LS increases to 11\,dB.

\begin{figure}[t!]
\vspace{-0.6cm}
\hspace{-3mm}
			\includegraphics[trim={0cm 0.cm 0cm 0cm},clip,width=3.6in]{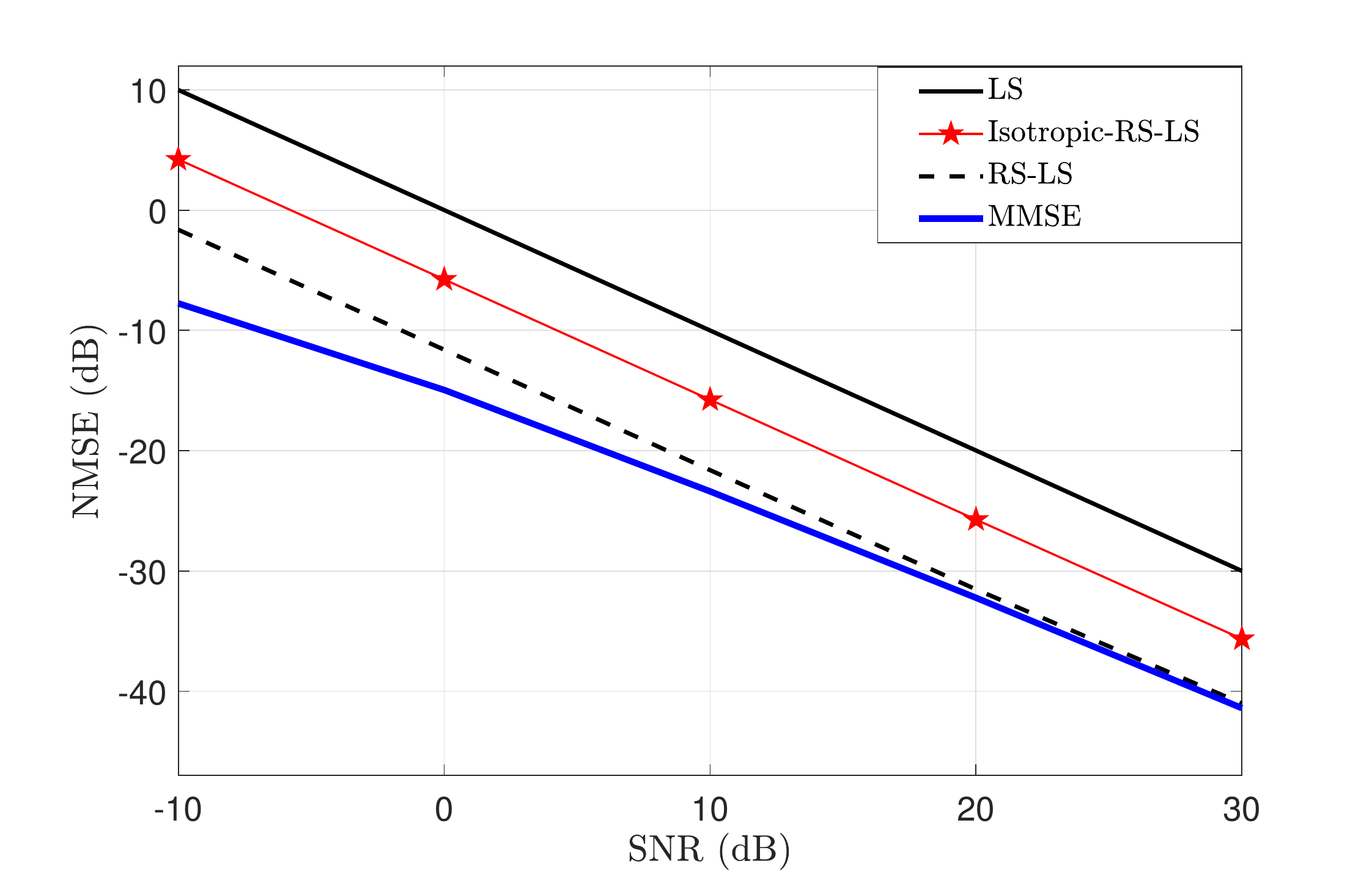}
			\vspace{-0.75cm}
			\caption{NMSE versus SNR for the $64\times64$ UPA  with $\Delta=\lambda/4$.} \label{fig:M64d25} \vspace{-5mm}
\end{figure}

When we keep the array size the same as in Fig.~\ref{fig:M128d125} but use less densely deployed antennas, i.e., $64\times 64$ UPA with $\Delta=\lambda/4$, we obtain the NMSEs shown in Fig.~\ref{fig:M64d25}. This figure has the same characteristics as Fig.~\ref{fig:M128d25}, where we have four times more antennas with same antenna spacing, i.e., $\Delta=\lambda/4$. This can be explained by the fact that the ratio of the rank of the spatial correlation matrix to $M$ is $\pi  (\Delta/\lambda)^2$ when $M$ is large and $\Delta$ is small \cite[Prop.~2]{Bjornson2021b}, which increases with $\Delta$ irrespective of $M$. Hence, when more antennas are deployed into a given array size as in Fig.~\ref{fig:M128d125}, the performance of the proposed estimator improves. Moreover, with more BS antennas in a given area, the capacity increases \cite[Fig.~2]{pizzo2020asilomar}.

\section{Conclusions}
We have derived a closed-form spatial correlation matrix expression for planar holographic massive MIMO arrays with clustered scattering and directive antennas. 
Since the antennas are densely deployed, the spatial correlation matrices become strongly rank-deficient, even with isotropic scattering.
We have proposed a novel channel estimation scheme that exploits the rank-deficiency induced by the array geometry without requiring the exact channel statistics. This estimator outperforms the conventional statistics-unaware LS estimator. The SNR gain is 10\,dB when 10\% of the eigenvalues are non-zero.
 
\section*{Appendix: Proof of Lemma~\ref{lemma:span}} 

We consider an arbitrary vector $\vect{x}\in \mathbb{C}^{M}$ that is in the nullspace of $\overline{\vect{R}}$.  If we can show that $\vect{x}$ is also in the nullspace of $\vect{R}$, we have completed the proof. We have 
 \begin{align} \label{eq:null-condition}
     \vect{x}^{\Htran} \overline{\vect{R}} \vect{x} = \beta  \iint_{\overline{\mathcal{F}}} \overline{f}(\varphi,\theta) \vert\vect{x}^{\Htran}\vect{a}(\varphi,\theta)\vert^2  d\theta d\varphi =0
 \end{align}
where we have used \eqref{eq:spatial-correlation} and $\overline{\mathcal{F}}$ is the domain of $\overline{f}(\varphi,\theta)$ where it is non-zero. If $\overline{f}(\varphi,\theta)$ includes some Dirac delta impulses, then we should have $\vect{x}^{\Htran}\vect{a}(\varphi,\theta)= 0$ at the corresponding angles $(\varphi,\theta)\in \overline{\mathcal{F}}$ to satisfy \eqref{eq:null-condition}. At the other angles where $\overline{f}(\varphi,\theta)$ is continuous, we first assume there is at least one pair of angles $(\varphi^{\star},\theta^{\star})\in \overline{\mathcal{F}}$ with $\vect{x}^{\Htran}\vect{a}(\varphi^{\star},\theta^{\star})\neq 0$. Then we can find a region of $ \overline{\mathcal{F}}$ around $(\varphi^{\star},\theta^{\star})$ such that   $\vect{x}^{\Htran}\vect{a}(\varphi,\theta)\neq 0$ from the continuity. This assumption results in a positive integral in \eqref{eq:null-condition}, which violates the initial assumption that the integral is zero. Hence, $\vect{x}$ should be orthogonal to $\vect{a}(\varphi,\theta)$, $\forall (\varphi,\theta)\in \overline{\mathcal{F}}$, and, hence $\forall (\varphi,\theta)\in {\mathcal{F}}$ where $\mathcal{F}$ is the domain of $f(\varphi,\theta)$ for which it is non-zero. This in turn leads to the fact that $\vect{x}^{\Htran}\vect{R}\vect{x}=0$ from \eqref{eq:spatial-correlation} and completes the proof.

\bibliographystyle{IEEEtran}
\bibliography{IEEEabrv,refs}

\end{document}